\documentclass[aps, prd, twocolumn, showpacs, superscriptaddress, groupedaddress]{revtex4-1} 
\usepackage{graphicx}	
\usepackage{amssymb}
\usepackage{dcolumn}
\usepackage{mathtools}
\usepackage{amsmath}
\usepackage{xcolor}
\usepackage{color}
\usepackage{subfigure, rotating, bm, array}
\usepackage[pagebackref=false, colorlinks=true]{hyperref}
\hypersetup{linkcolor=blue, citecolor=blue,
urlcolor=blue} 

\begin{document}

\title{Energy extraction from Janis-Newman-Winicour naked singularity}

\author{Vishva Patel}
\email{vishwapatel2550@gmail.com}
\affiliation{International Center for Cosmology, Charusat University, Anand-388421 (Gujarat), India}
\author{Kauntey Acharya}
\email{kaunteyacharya2000@gmail.com}
\affiliation{International Center for Cosmology, Charusat University, Anand-388421 (Gujarat), India}
\author{Parth Bambhaniya}
\email{grcollapse@gmail.com}
\affiliation{International Center for Cosmology, Charusat University, Anand-388421 (Gujarat), India}
\affiliation{PDPIAS, Charusat University, Anand-388421 (Gujarat), India}
\author{Pankaj S. Joshi}
\email{psjcosmos@gmail.com}
\affiliation{International Center for Cosmology, Charusat University, Anand-388421 (Gujarat), India}
\affiliation{International Centre for Space and Cosmology, School of Arts and Sciences,
Ahmedabad University, Ahmedabad-380009 (Gujarat), India}

\begin{abstract}
In general, energy extraction methods such as the Penrose process and the magnetic Penrose process are thought to be reliant on the existence of an ergoregion. Inside an ergoregion, there are negative energy states that allow a particle to extract energy and escape to an observer at infinity. In this paper, we considered the electromagnetic field in the rotating Janis-Newman-Winicour (JNW) spacetime. This concept is feasible because an accretion disc forms an electromagnetic field around compact objects. After that, we briefly examine negative energy orbits and their significance in energy extraction. The ergoregion is absent in a rotating JNW geometry, but we show that the effective ergoregion is there. The change in a negative energy orbit concerning the magnetic field (B), spin parameter (a), and electric charge (Q) is analyzed. We find that the total energy extraction efficiency within this process can be around $60\%$ for the rotating JNW naked singularity.

\end{abstract}
\maketitle

\section{Introduction}
\label{sec_intro}

The major concerns in high-energy astrophysics these days are the powering of active galactic nuclei, X-ray binaries, and quasars. To understand the reason behind these high-energy phenomena such as Gamma Ray Bursts, Fast Radio Bursts, and the formation of jets, many efforts have been made by researchers by studying the energy extraction mechanisms. These mechanisms have been studied for various compact objects, but mainly for black holes. On the other hand, it is well known now that by the end state of collapsing star both a visible and hidden singularity can form \cite{joshi,goswami,mosani1,mosani2,mosani3,mosani4,Deshingkar:1998ge,Jhingan:2014gpa,Joshi:2011zm}. On that count, the presence or absence of the event horizon should be responsible for the distinct physical signature of a black hole and naked singularity. A number of articles in the literature are devoted to investigating accretion disks, tidal disruption, gravitational lensing, and timelike and null geodesics around such compact objects \cite{Page:1974he,Liu:2020vkh,Liu:2021yev,Joshi:2013dva,Bambhaniya:2021ugr,Tahelyani:2022uxw,Rahaman:2021kge,Harko:2008vy,Harko:2009xf,Kovacs:2010xm,Guo:2020tgv,Chowdhury:2011aa,Lattimer:1976kbf,Chicone:2004pv,Madan:2022spd,Shaikh:2019jfr,Shaikh:2018oul,Paul:2020ufc,Virbhadra:2007kw,Gyulchev:2008ff,Kala:2020prt,Sahu:2012er,Martinez,Madan:2022spd, Eva1, Eva2, tsirulev, Joshi:2019rdo, Bambhaniya:2019pbr, Dey:2019fpv, Bam2020,Bambhaniya:2022xbz,Gralla:2019xty,Vagnozzi:2019apd,Chen:2022nbb,Dey:2020haf,atamurotov_2015,abdujabbarov_2015b,Li:2021,Hu:2020usx,Patel:2022vlu,Kaur:2021wgy,Saurabh:2020zqg,Pugliese:2022oes,Vagnozzi:2022moj,Saurabh:2022jjv}. As expected, studying these properties indeed gives a distinct observable signature. In addition to that, for the energy extraction due to the absence of an event horizon, naked singularity should give high energy extraction efficiency rate than the black hole. \\

The exciting idea of energy extraction leads back to 1969 when R. Penrose and R. M. Floyd gave the rotational energy extraction mechanism \cite{Penrose:1971uk}, in which they considered the ergoregion of Kerr black hole where negative energy orbits exist. Later on, numerous studies were carried out where the effects of the magnetic field around the compact objects on the energy extraction processes were considered which is known as the magnetic Penrose process. Apart from this, the concept of collision of particles in the Kerr black hole was also developed, now known as the collisional Penrose process \cite{Wagh:1985vuj,Tursunov:2019oiq,Wagh:1989zqa,Patil:2011yb,Patil:2014lea,Patil:2015fua,Hejda:2021cbk,Khan:2019gco,Okabayashi:2019wjs,Mukherjee:2018cbu,Schnittman:2018ccg,Tanatarov:2016mcs}.\, The radiative Penrose effect and its potential significance to the formation of relativistic jets are discussed in \cite{Kolos:2020gdc}.\, The general relativistic rotational energy extraction from the accretion disk has been investigated in \cite{Pugliese:2021ivl}.\, A few other mechanisms of energy extraction such as the Blandford-Znajek (BZ) Process, and the Blandford Payne process, involving rotating black holes as a source, also extensively include the effects of magnetohydrodynamics \cite{Bz}. There are several other processes where the collision of particles in the vicinity of static and rotating compact objects have also been studied which are the Banados-Silk-West (BSW) process and particle acceleration process \cite{Banados:2009pr,
Zaslavskii:2012fh,Zaslavskii:2012yp,Zaslavskii:2012ax,Pradhan:2016pgn,Zaslavskii:2020agl,Zaslavskii:2022fwv,Komissarov:2008yh,Konoplya:2021qll,Kimura:2021dsa}.\\
 
In addition to this, many articles show these energy extraction mechanisms for other classes of compact objects as well along with Kerr black hole. The Magnetic Penrose process is studied for the Kerr naked singularity in which energy extraction efficiency is about 150\% \cite{Kerrnakedsingularity}. The energy extraction using the Super Penrose process is also studied in which Reissner–Nordström (RN) naked singularity is considered as a central compact object \cite{Zaslavskii:2022nbm}. Authors have explored particle acceleration and energy extraction for singular as well as regular black hole spacetimes \cite{Patil:2012fu,Nakao:2013uj,Patel:2022jbk}. However, mechanisms related to rotational energy extraction have not been explored for naked singularity spacetimes as compared to black hole spacetimes. Therefore in this paper, we will analyze the presence of an ergoregion in the naked singularity spacetime. \\
 
There are multiple studies investigating the behaviour of an ergoregion and negative energy orbits \cite{Chakraborty:2016ipk,Pani:2010jz,Zaslavsky:2013dra}.\, For the JNW naked singularity spacetime, many properties including an ergoregion have been studied in \cite{Karmakar:2017lho}, where the authors show that an ergoregion does not exist in rotating JNW naked singularity spacetime. In this paper, we consider the rotating JNW naked singularity spacetime in the electromagnetic field. This configuration can act as a good source that emits highly energetic radiation. However, we show that energy extraction mechanisms such as the Penrose process and other rotational energy extraction mechanisms defined by modifying the Penrose process, do not require the existence of an ergoregion, rather it mainly depends on the behaviour of negative energy orbits. Another article also suggests this fact \cite{Denardo:1973pyo}. When a rotating compact object in the magnetic field is considered, particles can extract energy from the rotation of the geometry as well as the magnetic field around the compact object.\\
 
In this paper, the effect of the magnetic field around the rotating JNW naked singularity spacetime is studied. This concept of considering the magnetic field around the compact object is certainly interesting, but not surprising. A magnetic field appears in very small-scale stars for example the Sun. Thus studies also include magnetic fields around black hole spacetimes as it can give answers to many questions, mostly related to the high energy phenomena as one can consider energy released through excitation of charged particles along the magnetic field lines, as well as phenomena such as magnetic re-connection (MR) \cite{Stuchlik:2019dlx}.  \\ 
 
This paper is organised as follows: we discuss the JNW naked singularity metric and electromagnetic four components of the external electromagnetic field in Section (\ref{sec_jnwgeom}). In section (\ref{sec_neo}), we first explain the importance of negative energy orbits and then analyze the behavior of negative energy orbits for rotating JNW spacetime with the uniform electromagnetic field. In section (\ref{sec_energyextra}), we discuss the magnetic Penrose process and the behaviour of the energy extraction efficiency with magnetic field parameter $B$, the spin parameter $a$, and JNW metric parameter $\nu$. In the last section (\ref{sec_conclusion}), we discuss the results and conclusions of this study. Throughout the paper, we have considered geometrized units. Hence, the gravitational constant (G) and the speed of light (c) are set equal to one. The signature of the metric is considered as (-,+,+,+).

\section{Janis-Newman-Winicour naked singularity spacetime}
\label{sec_jnwgeom}

The extended geometry of Schwarzschild spacetime with a massless scalar field is known as Janis-Newman-Winicour spacetime. The rotating version of the JNW metric is derived in \cite{Solanki:2021mkt}, using Newman-Janis Algorithm (NJA) without complexification. The rotating JNW naked singularity spacetime can be written as,
\begin{eqnarray}
    ds^2 = -\left(1-\frac{2f}{\rho^2} \right)dt^2 - \frac{4af\sin^2\theta}{\rho^2} dt d\phi  + \frac{\Sigma\sin^2\theta}{\rho^2} d\phi^2 \nonumber \\   + \frac{\rho^2}{\Delta} dr^2 + \rho^2 d\theta^2,
    \,\,\,\,\,\,\,\,\,\,
    \label{rotatingjnwmetric}
\end{eqnarray}
where,
\begin{eqnarray}
    && \nu\,=\,\frac{2\,m}{b}, \label{nu} \\
    && b = 2\,\sqrt{m^{2}\,+\,q_{s}^{2}}, \label{b}\\
    && f\,=\,\frac{1}{2} r^2 \left(\left(1-\frac{2 m}{\nu  r}\right)^{-\nu }-1\right) \left(1-\frac{2 m}{\nu  r}\right),\\
    && \rho^{2}\,=\,a^2 \cos ^2(\theta )+r^2 \left(1-\frac{2 m}{\nu  r}\right)^{1-\nu },\\
    && \Sigma\,=\,(a^2 \sin ^2(\theta )+\rho^{2})^2-a^2 \sin ^2(\theta ) \Delta.\\
    && \Delta = a^2-\frac{2 m r}{\nu }+r^2,
\end{eqnarray}   
where $m$, $a$, and $q_{s}$ represent the ADM mass of spacetime, spin parameter, and scalar field charge respectively. In the spacetime, the limit for the $\nu$ is $0 < \nu < 1$ and the scalar field charge is non-zero. The Eq.\,(\ref{rotatingjnwmetric}) is reduced to a Kerr black hole and static JNW naked singularity when $\nu\,=\,1$ and $a\,=\,0$ respectively.\\

There are multiple methods to consider the different profiles of the magnetic field around the compact object as the exact distribution of the magnetic field is unknown. Here, we consider an external magnetic field with intensity B which is asymptotically uniform  \cite{Stuchlik:2019dlx}. The field lines are oriented along the z-axis, i.e., orthogonal to the equatorial plane of the spacetime geometry. This can be described using the electromagnetic four components $A_{\mu}$ having two nonzero components of the form,

\begin{eqnarray}
    &&A_{t}=\frac{B}{2}(g_{t\phi} + 2 a g_{tt})-\frac{Q}{2}g_{tt}-\frac{Q}{2},\label{At}\\
    && A_{\phi}=\frac{B}{2}(g_{\phi\phi} + 2 a g_{t \phi})-\frac{Q}{2}g_{tt},\label{aphi}
\end{eqnarray}
where $Q$ is the externally induced charge in the rotating JNW geometry. In the next section, we derive the equations of motion for the charged particle to observe the behavior of negative energy orbits in the rotating JNW geometry, in an external asymptotically uniform magnetic field $B$.

\section{Negative energy orbits (NEO)}
\label{sec_neo}

 Penrose and Floyd showed that how particles can extract the energy from an ergoregion of a Kerr black hole, where the negative energy orbits are also possible. Later on, Wheeler suggested that the Penrose process can act as a possible mechanism to explain the high energy phenomena in the universe. Thus many research works are carried out in the domain of rotational energy extraction and discuss the ergoregion of compact objects \cite{Pani:2010jz,Chakraborty:2016ipk,Zaslavsky:2013dra}. However, studies suggest that energy extraction requires the existence of negative energy orbits \cite{Denardo:1973pyo}. The ergoregion is not always necessary for energy extraction phenomena. Therefore in this section, we first discuss the negative energy orbits.\\

Negative energy orbits are energy states in which the total energy of the particle is negative with respect to an observer at infinity. It can be easily understood for classical mechanics, where one can consider a system of a central object and a particle. If the particle is far enough from the central object, the potential energy of a particle far from the central object is taken to be zero, as the point of zero potential energy can be chosen arbitrarily. Thus if the particle is near the central object with zero kinetic energy, the total energy of the particle can be considered to be negative. If the particle gets the energy equivalent to the negative energy, it will escape to the infinite distance where it possesses the zero potential energy i.e. it will be free from the gravitational influence of the central object.\\

\begin{figure*}[ht!]
\centering
\subfigure[ $\nu$ = 0.3, a = 0.9, Q = 0.5, $L$ = -13.]
{\includegraphics[width=5.5cm]{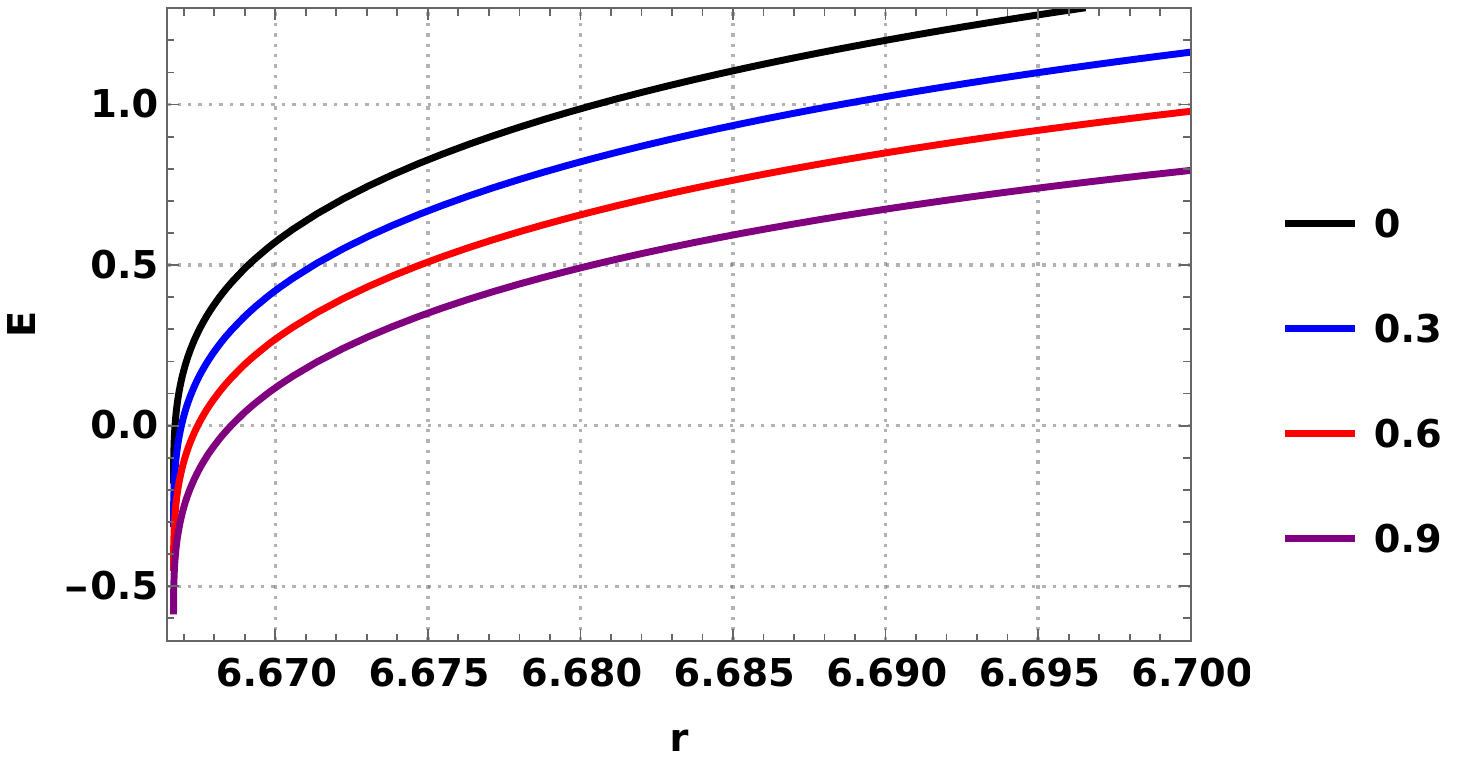}\label{fig:11}}
\hspace{0.4cm}
\subfigure[ $\nu$ = 0.3, a = 0.9, Q = 0.9, $L$ = -13.]
{\includegraphics[width=5.5cm]{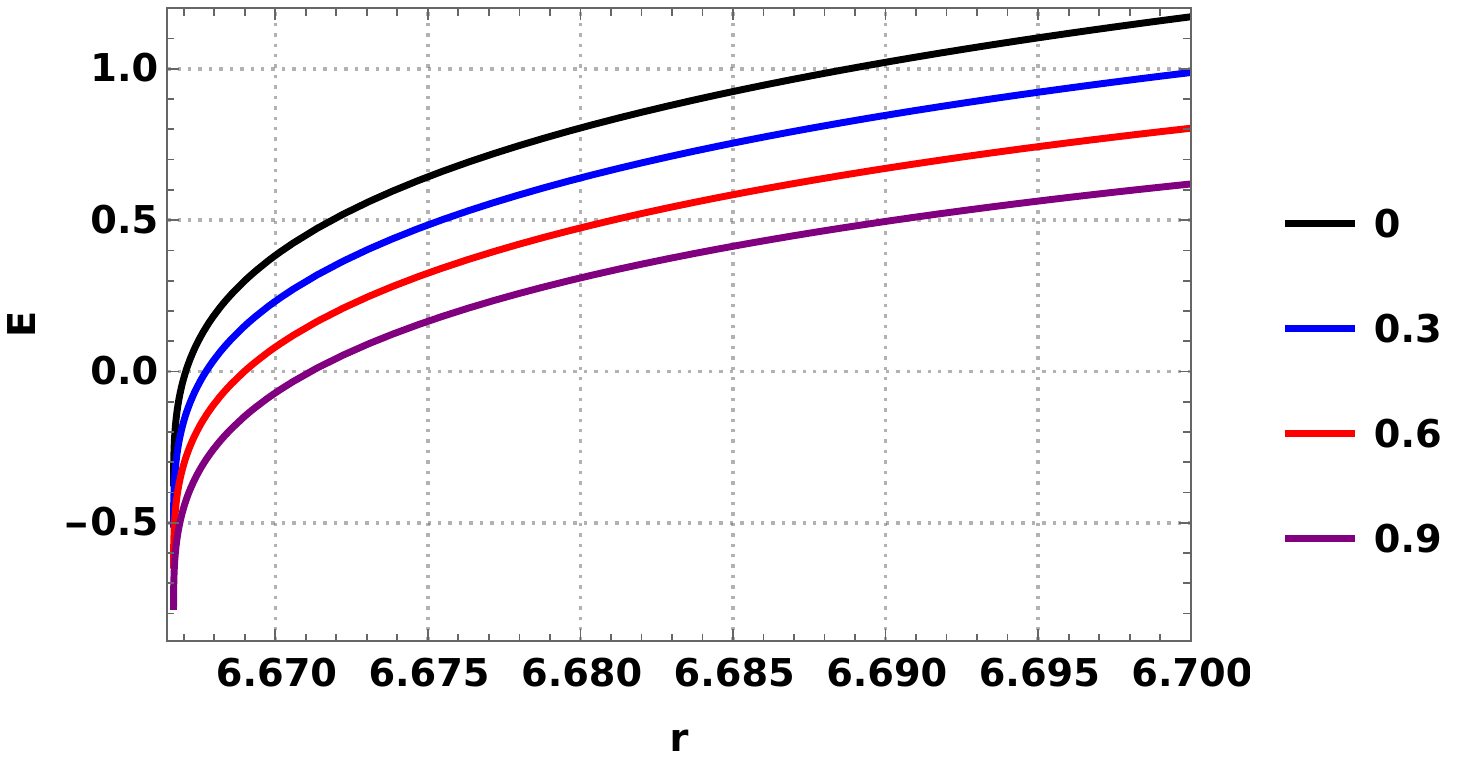}\label{fig:12}}
\hspace{0.4cm}
\subfigure[ $\nu$ = 0.5, a = 0.1, Q = 0.1, $L$ = -3.]
{\includegraphics[width=5.5cm]{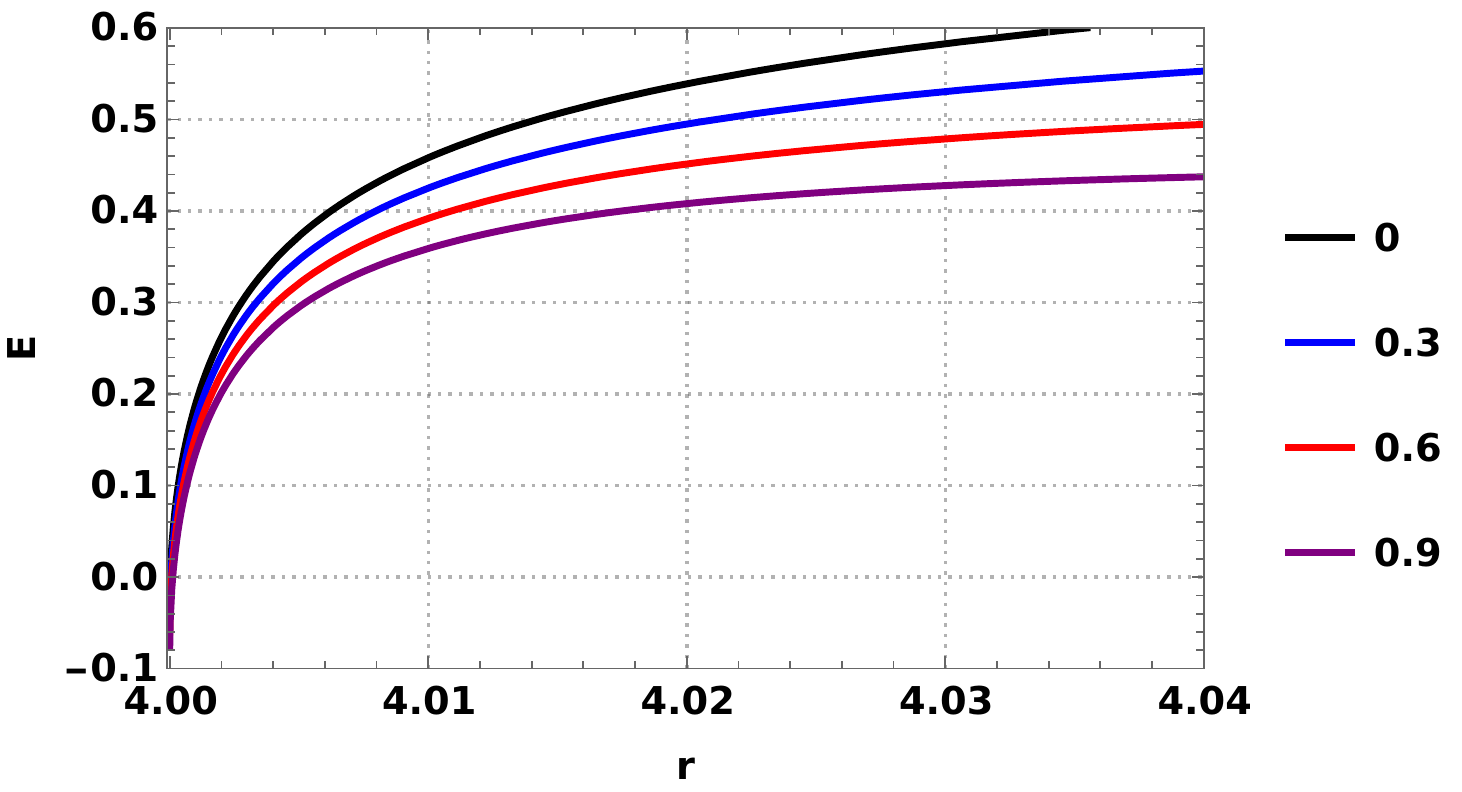}\label{fig:13}}
\hspace{0.4cm}
\subfigure[$\nu$ = 0.5, a = 0.9, Q = 0, $L$ = -20.]
{\includegraphics[width=5.5cm]{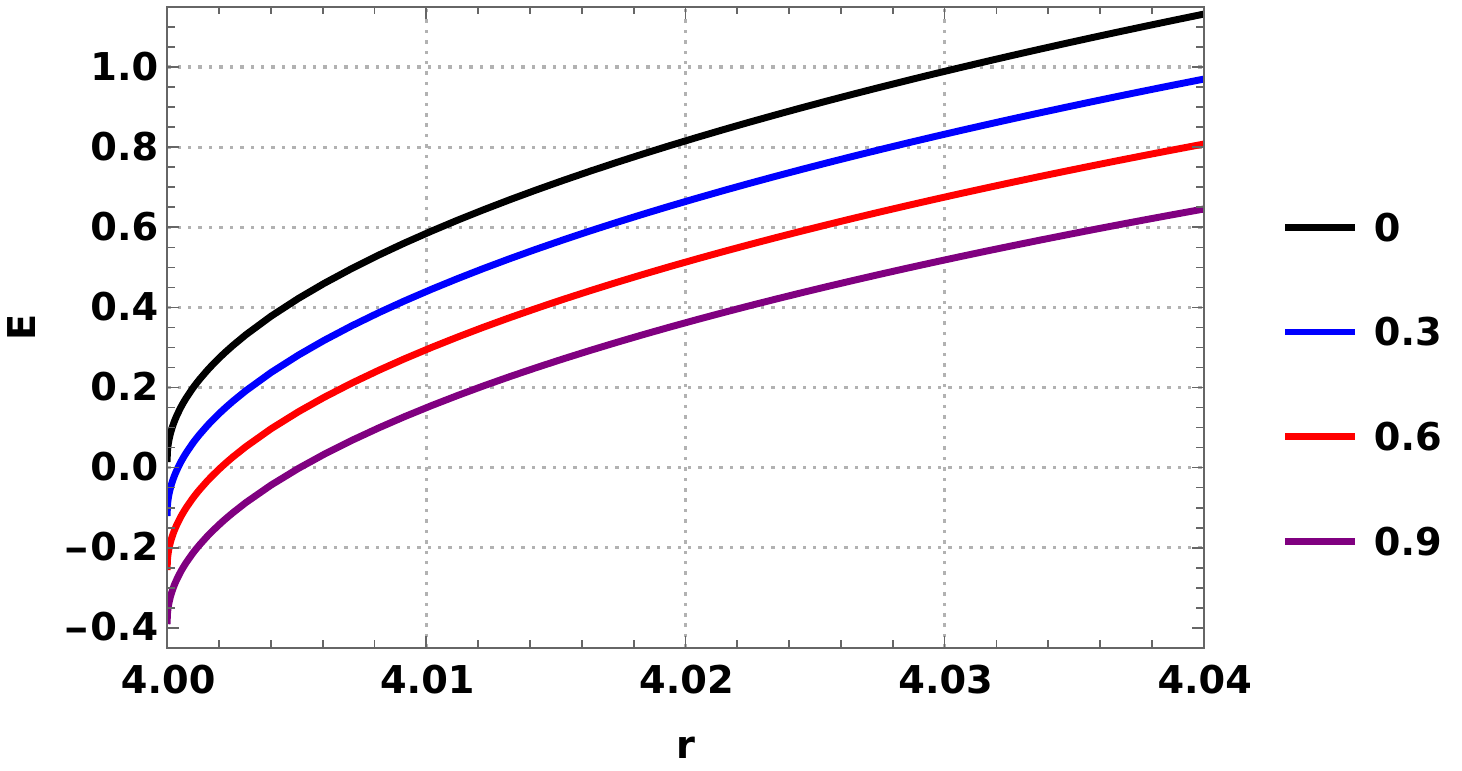}\label{fig:14}}
\hspace{0.4cm}
\subfigure[$\nu$ = 0.5, a = 0.9, Q = 0.5, $L$ = -13. ]
{\includegraphics[width=5.5cm]{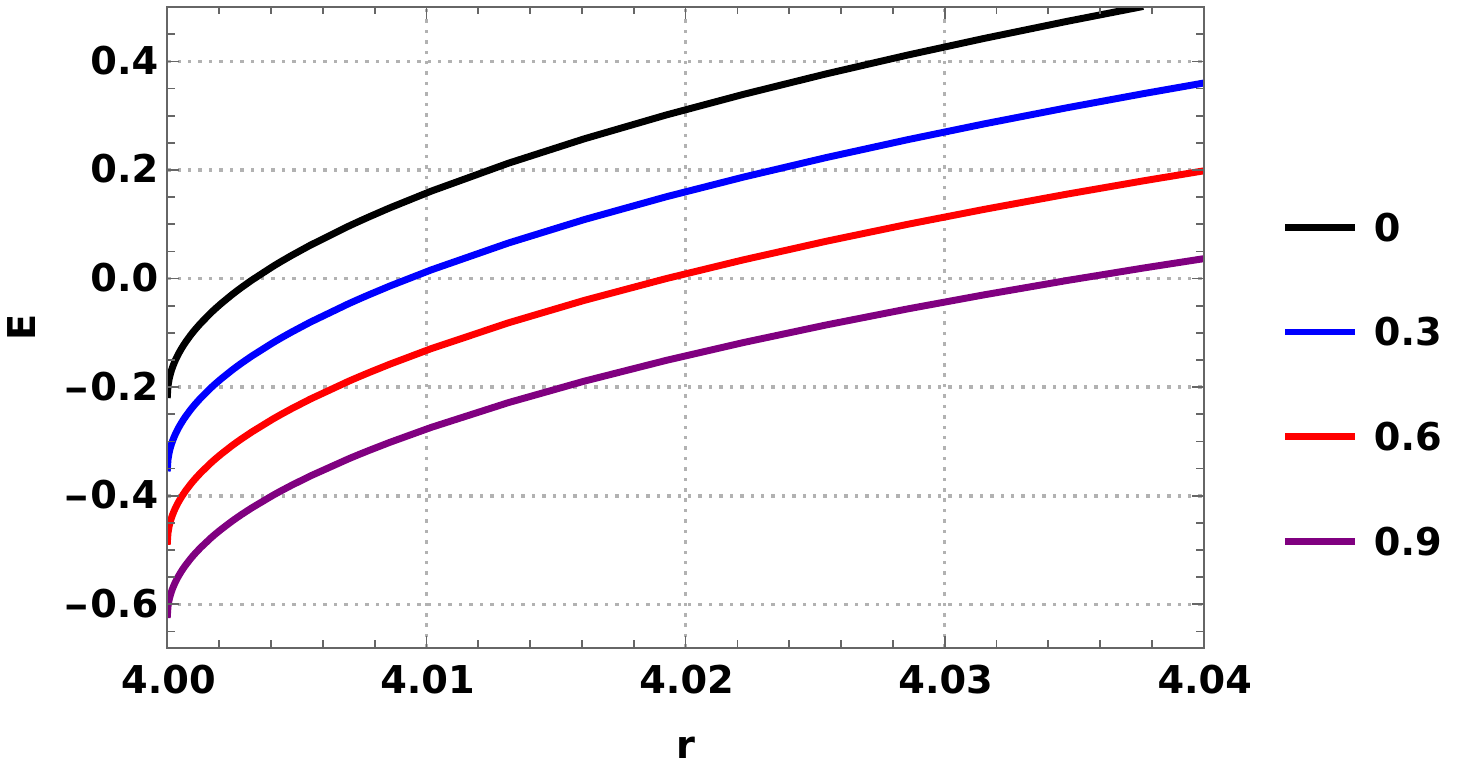}\label{fig:15}}
\hspace{0.4cm}
\subfigure[$\nu$ = 0.5, a = 0.9, Q = 0.9, $L$ = -13. ]
{\includegraphics[width=5.5cm]{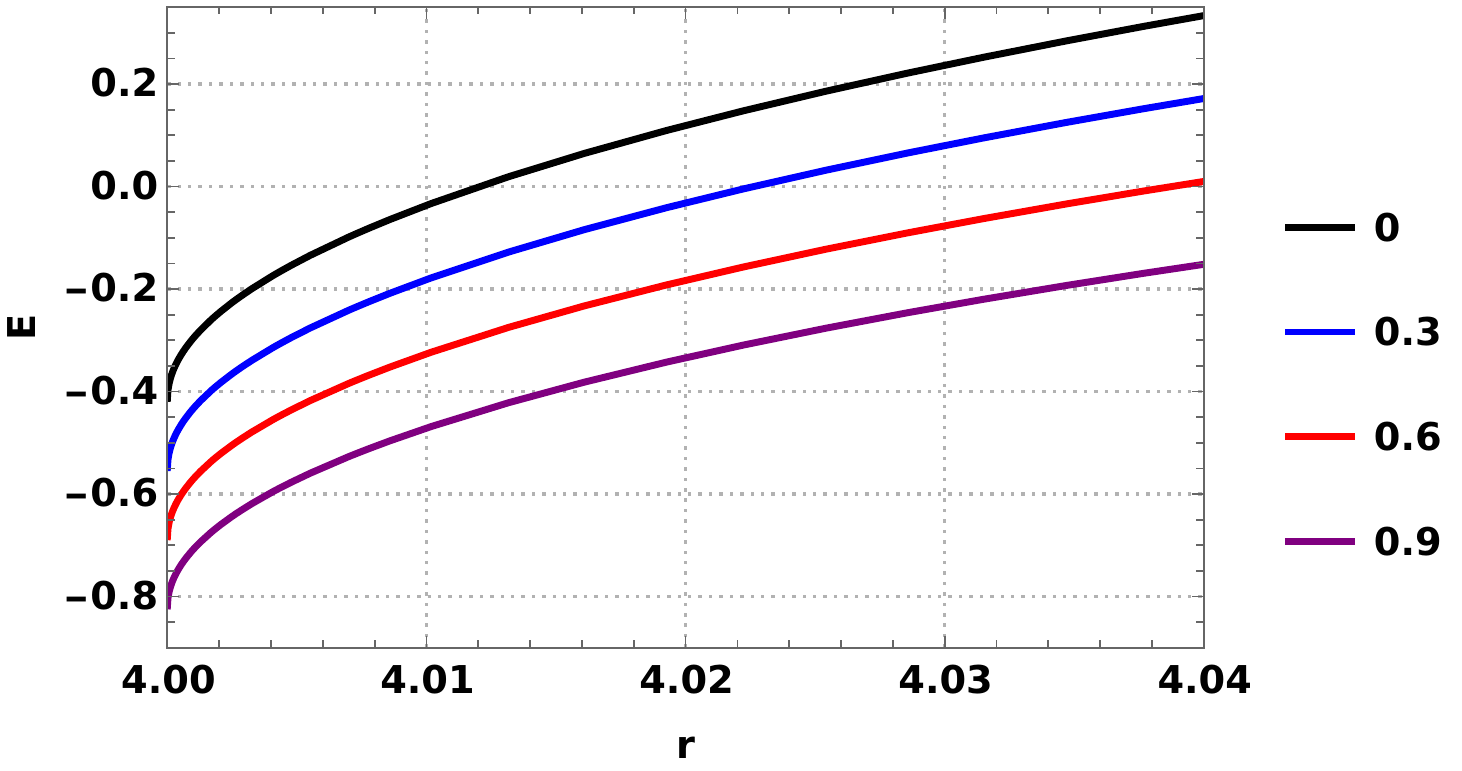}\label{fig:16}}
\hspace{0.4cm}
\subfigure[$\nu$ = 0.7, a = 0.9, Q = 0, $L$ = -20. ]
{\includegraphics[width=5.5cm]{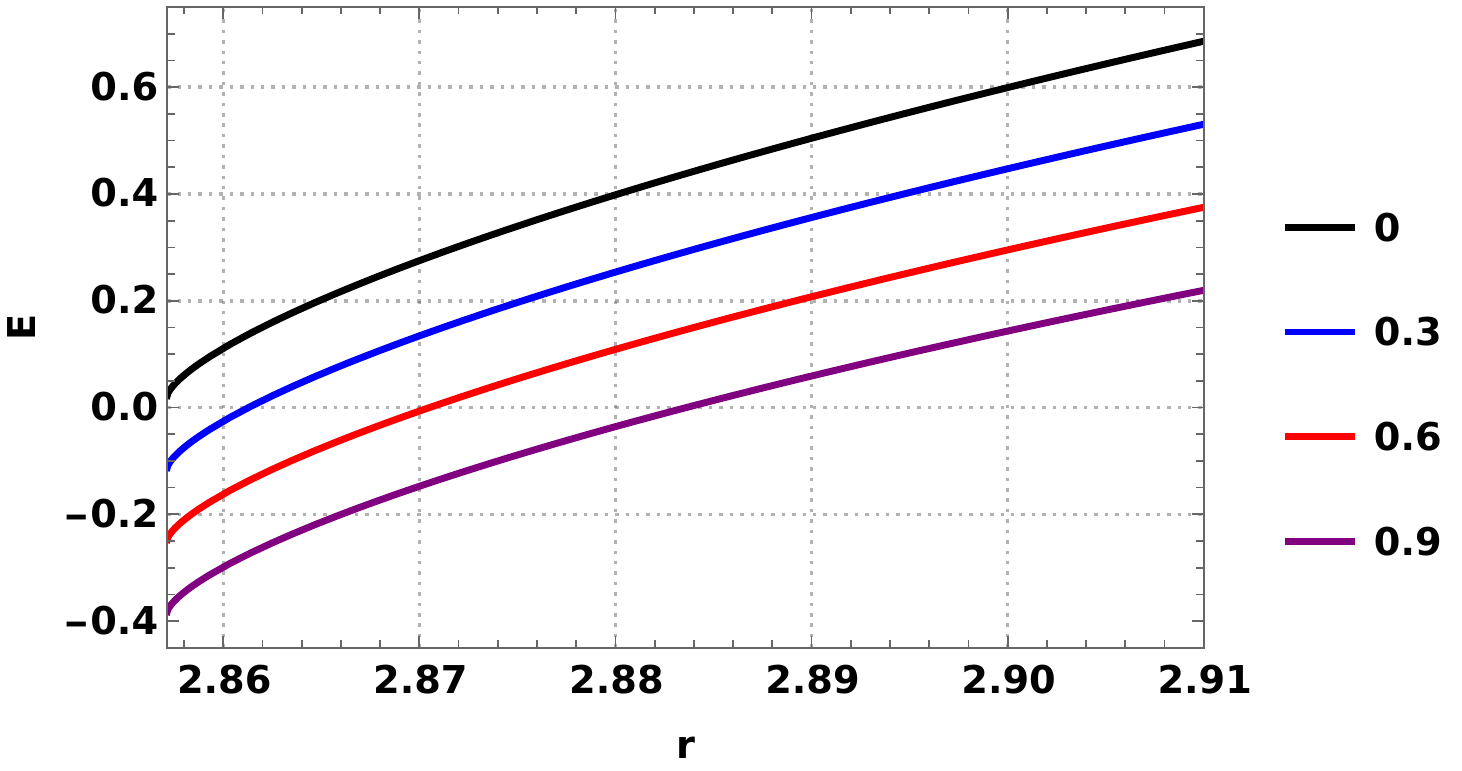}\label{fig:17}}
\hspace{0.4cm}
\subfigure[$\nu$ = 0.7, a = 0.9, Q = 0.5, $L$ = -15. ]
{\includegraphics[width=5.5cm]{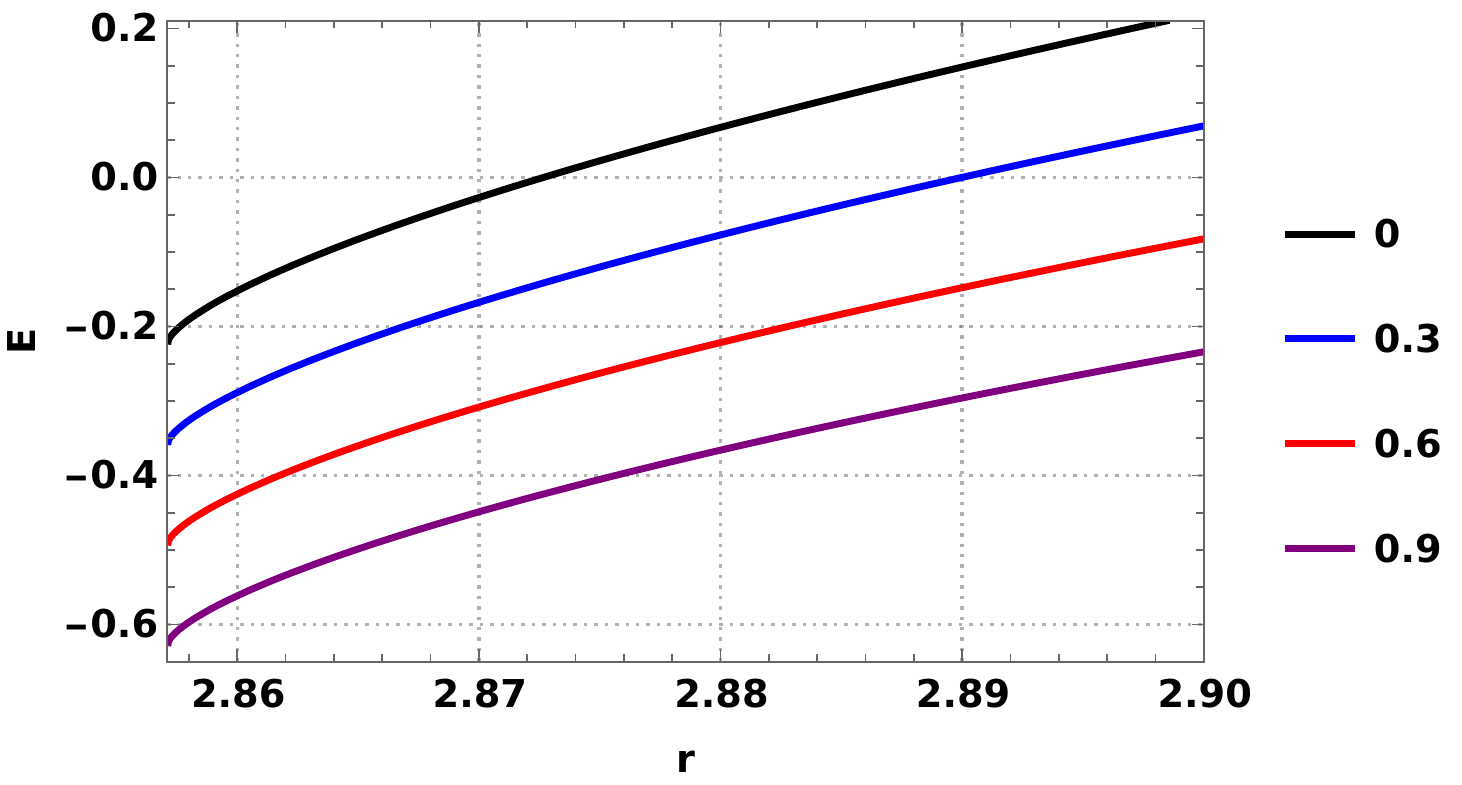}\label{fig:18}}
\hspace{0.4cm}
\subfigure[$\nu$ = 0.7, a = 0.9, Q = 0.9, $L$ = -15. ]
{\includegraphics[width=5.5cm]{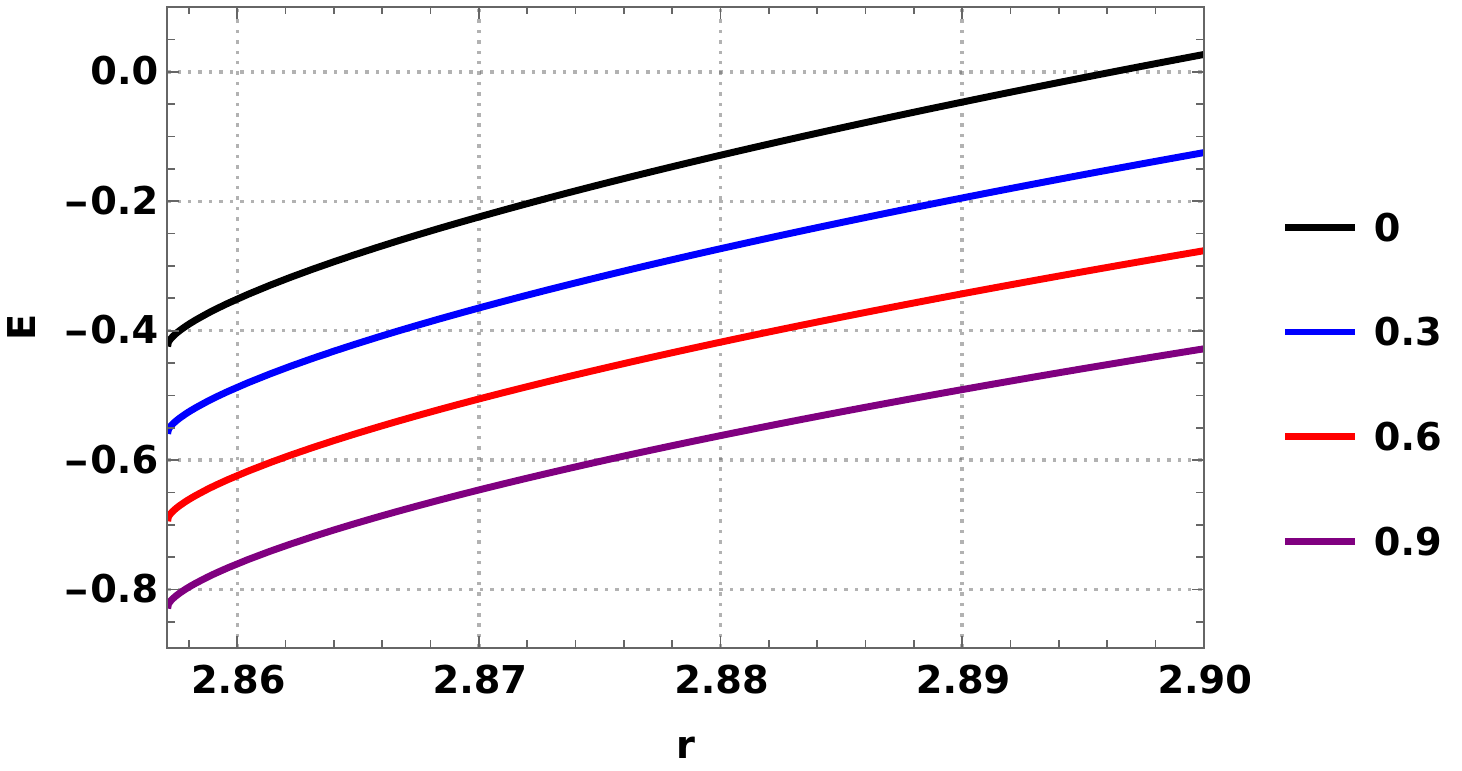}\label{fig:19}}
\hspace{0.4cm}
\subfigure[$\nu$ = 0.9, a = 0.9, Q = 0, $L$ = -20. ]
{\includegraphics[width=5.5cm]{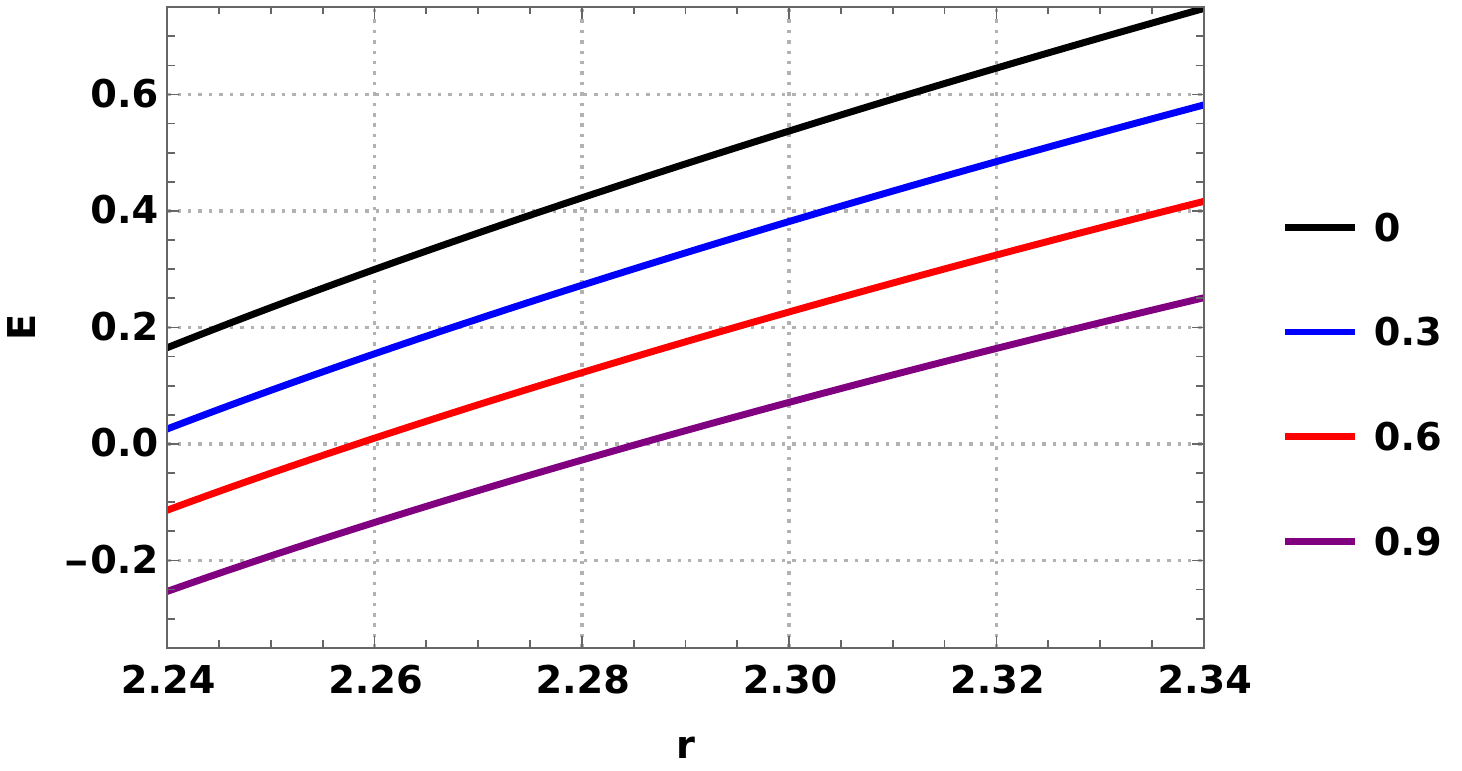}\label{fig:110}}
\hspace{0.4cm}
\subfigure[$\nu$ = 0.9, a = 0.9, Q = 0.5, $L$ = -20. ]
{\includegraphics[width=5.5cm]{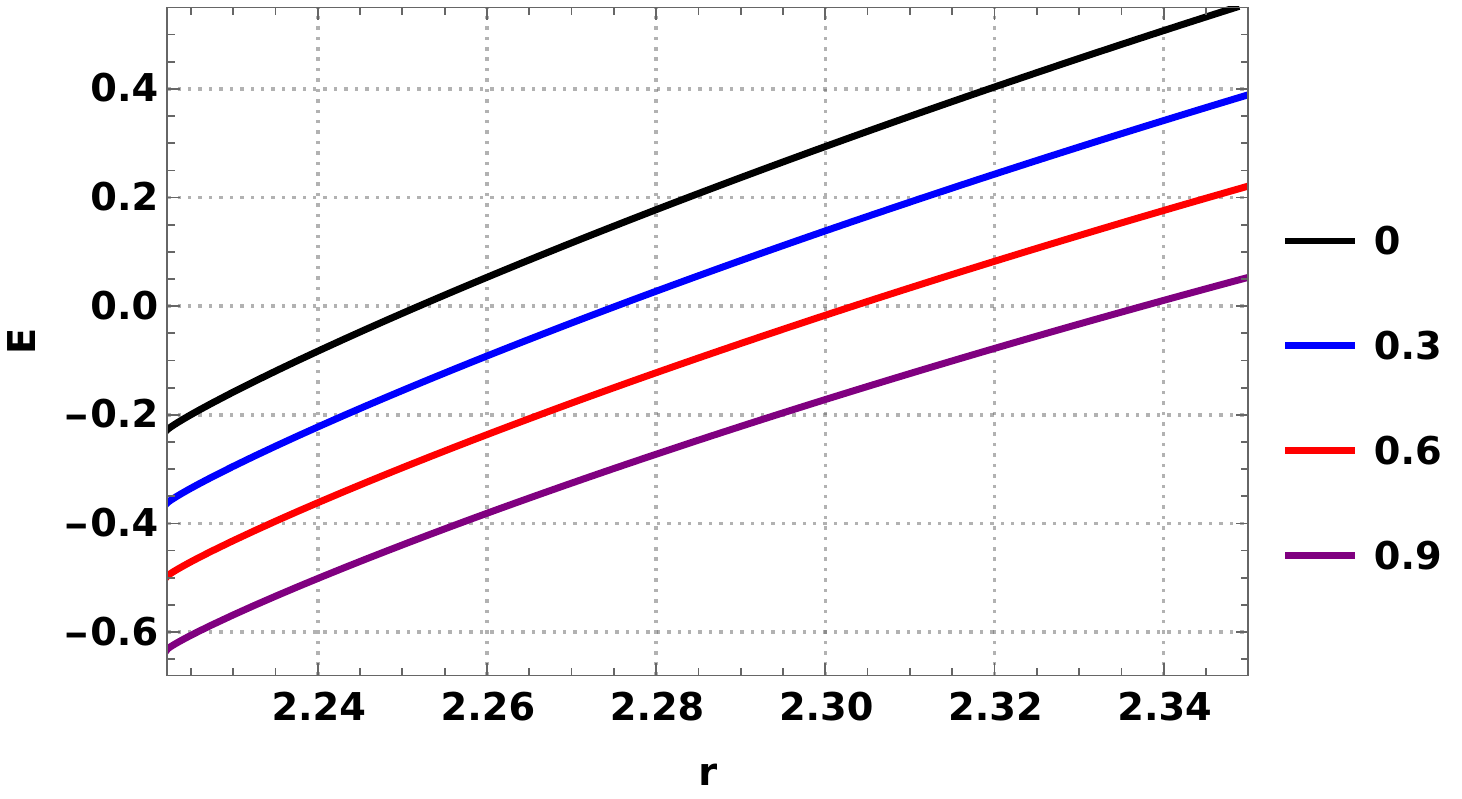}\label{fig:111}}
\hspace{0.4cm}
\subfigure[$\nu$ = 0.9, a = 0.9, Q = 0.9, $L$ = -20. ]
{\includegraphics[width=5.5cm]{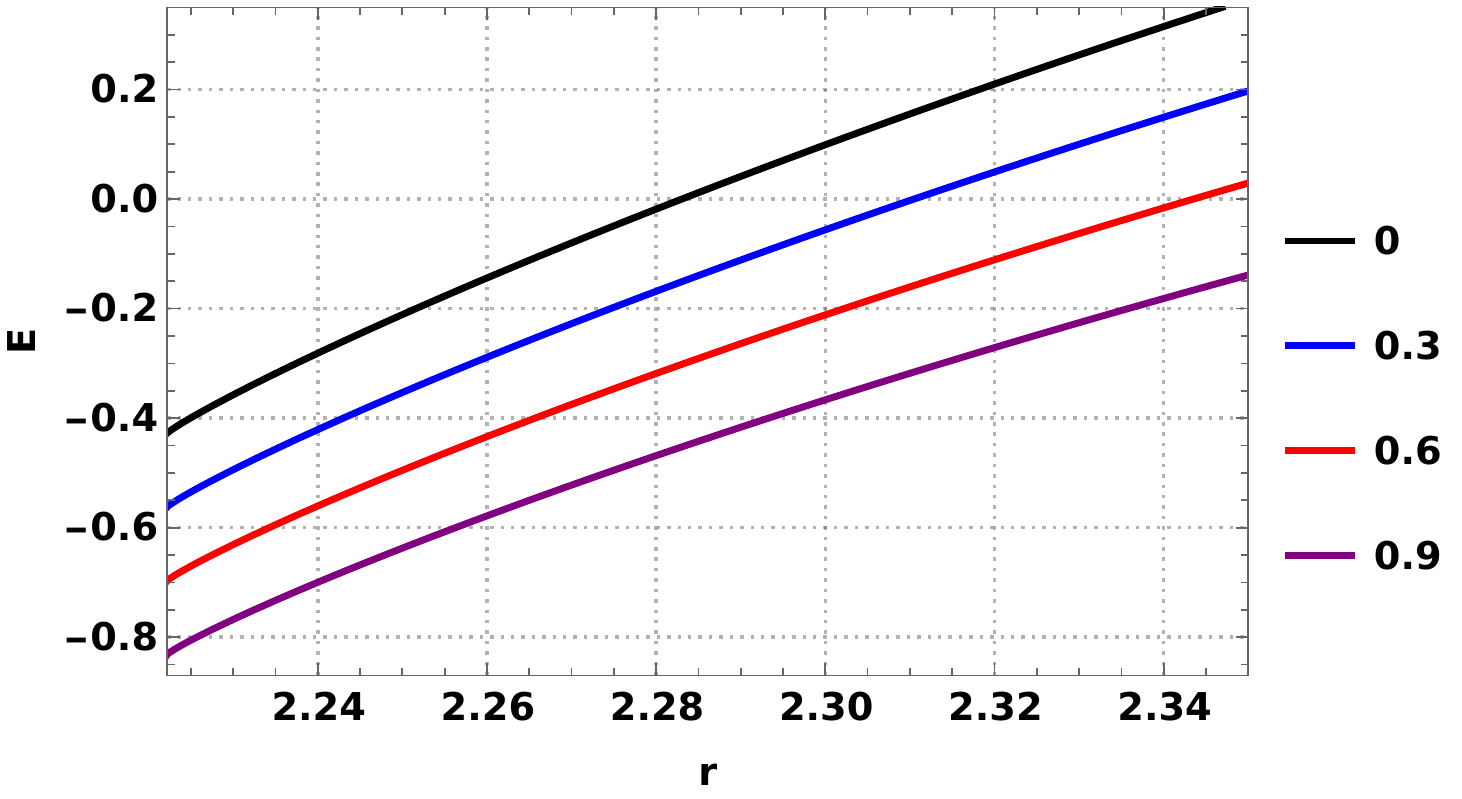}\label{fig:112}}
\hspace{0.4cm}
 \caption{The above figures show the change in the total energy with respect to the radial distance in the rotating JNW geometry. The different values on the right side of the plots indicate the intensity of the magnetic field (B) around the compact object. The mass of the object is taken as unity for all different values of $\nu$.}
 \label{fig:1}
\end{figure*}

However, in special relativity, a timelike particle always possesses rest mass energy. Even if the particle is far enough from any central object, it will have non-zero rest mass energy. If the particle is near a Schwarzschild black hole, it needs energy equivalent to the rest mass energy to escape the gravitational influence of the Schwarzschild black hole. But in the case of the Kerr black hole, when the particle is inside the ergoregion, the particle possesses rest mass energy as well as energy because of the frame-dragging effect. In addition to that, presence of electromagnetic field also increases the energy of the particle near the central compact object. In this case, the particle require energy equivalent to the sum of its rest mass energy and energy because of other properties such as rotation and electromagnetic field to escape from the gravitational influence of the central compact object. This energy state of the particle is known as the negative energy state. Since the particle possesses the negative energy state only in certain orbits, those orbits are known as negative energy orbits. For the Kerr black hole, the negative energy orbits exist only inside the ergoregion.\\

In the Penrose process, a particle splits into two particles inside an ergoregion of the Kerr black hole. One particle enters into an orbit with a negative energy state and falls inside the central black hole, while the other particle gets the energy equivalent to the frame-dragging energy of the particle, in addition to the rest mass energy and thus escapes to infinity. As mentioned for Kerr black hole case, the negative energy orbits exist only inside the ergoregion, thus Penrose process is defined only inside the ergoregion.\\

Following this, one can understand that a particle can possess a negative energy state because of certain properties of the compact object such as rotation, and electromagnetic. Other particle can extract this energy and escape to infinity. Thus in this section, we first discuss the negative energy orbits of a particle around a rotating compact object in the presence of a general electromagnetic field tensor $A_{\mu}$. We then discuss the negative energy orbits for a charged particle around a rotating JNW naked singularity in an asymptotically uniform magnetic field. \\

Consider a rotating compact object metric $g_{\mu \nu}$ surrounded with an electromagnetic field denoted by general electromagnetic four components $A_{\mu}$. For the simplicity, we consider that a particle with charge $q$ and non-zero mass $m$ is moving in an equatorial plane. The conserved energy and angular momentum can be written as,
\begin{eqnarray}
    && - E\,=\,g_{tt} \dot{t} + g_{t\phi} \dot{\phi} + q\,A_{t}, \label{e}\\
    && L\,=\, g_{\phi \phi} \dot{\phi} + g_{t\phi} \dot{t} + q\,A_{\phi} \label{l}.
\end{eqnarray}
From Eqs.\,(\ref{e}) and (\ref{l}), expressions of four velocity of the charged particle $u^{t}$ and $u^{\phi}$ can be written as,

\begin{eqnarray}
    && u^{\phi} = - \left( \frac{E g_{t \phi}+\,L\, g_{tt}}{g_{t \phi}^{2}- g_{tt} g_{\phi \phi}}\right)- q \left( \frac{A_{t} g_{t \phi} - A_{\phi} g_{tt}}{g_{t \phi}^{2}- g_{tt} g_{\phi \phi}}\right) \label{phidot},\\
    && u^{t} = - \left( \frac{E g_{\phi \phi}+\,L\, g_{t \phi}}{g_{tt} g_{\phi \phi}-g_{t \phi}^{2}}\right)- q \left( \frac{A_{t} g_{\phi \phi} - A_{\phi} g_{t \phi}}{g_{tt} g_{\phi \phi}-g_{t \phi}^{2}}\right) \label{tdot}.
\end{eqnarray}
From $u_{\alpha}\,u^{\alpha} = -1$ we can write the equation of total energy for the motion of a charged particle in an equatorial plane at any constant radius ($\theta\,=\, \pi/2$),
\begin{equation}
   g_{tt}\,(u^{t})^{2}\,+\,2\,g_{t\phi}\,(u^{t})\,(u^{\phi})\,+\,g_{\phi \phi}\,(u^{\phi})^{2}\,=\,-1. \label{geode}
\end{equation}
Using Eq.\,(\ref{phidot}), and (\ref{tdot}) and solving it for $E$, we get general equations that describe the energy states of the particle in orbits around the compact object. \\

Using the Eq.\,(\ref{rotatingjnwmetric}) and expressions of four components of the electromagnetic four vectors from Eq.\,(\ref{At}) as well as Eq.\,(\ref{aphi}) in Eq.\,(\ref{geode}) to obtain the expression for energy states of the charged particle around the rotating JNW naked singularity, which is given in the Appendix (\ref{sec_app}).\\

Here, we consider the negative energy orbits $E\,<\,0$ for $L\,<\,0$, and the charge of a moving particle and the charge of the electromagnetic four components are such that the condition $q\, Q\,<\,0$ is maintained. The reason behind taking $L\,<\,0$ is, in the rotating compact object two scenarios are considered for the motion of a particle. In the first scenario, a particle is moving closer to the compact object in the same direction as the object's spin ($L\,>\,0$). The second one is, a particle coming from the opposite side of the objects' spin ($L\,<\,0$). Hence, a particle coming from the same direction as the object's spin will eventually move in the object's direction. While the particle coming from the opposite direction, have to experience the frame-dragging effect and after that, it also goes in the direction of the object's spin. In a static case, due to the absence of a frame-dragging effect the results with $L\,<\,0$ and $L\,>\,0$ are similar. The other condition with charge is because we considered the electric charge in the geometry. As per the nature of an electric charge, having the same charges will turn into a repulsive effect. To avoid that the $q\, Q\,<\,0$ condition is considered. \\

In Fig.\,(\ref{fig:1}), the negative energy states in the static JNW geometry are shown for different values of the metric parameter $\nu$, magnetic field parameter $B$, induced charge $Q$, and spin parameter $a$. It is clear that the energy state of the particle is negative before a certain radial distance for given values spin parameter $a$, magnetic field intensity parameter $B$, and the metric parameter $\nu>0.1$. As we increase the values of either of the parameter among metric parameter $\nu$, the spin parameter $a$, and magnetic field parameter $B$, keeping the values of the other parameters constant, the region of negative energy orbits increases. As discussed above about negative energy orbit, we can expect that energy extraction efficiency would increase with this increasing behavior of negative energy orbits as the particle will get more energy to escape to infinity. This expected behavior of energy extraction efficiency is also observed in the next section. In the range of $ 0.5\,<\,\nu\,<\,1 $, the negative energy orbits exist for every range of angular momentum.  The detailed discussion regarding Fig.\,(\ref{fig:1}) is given in section (\ref{sec_conclusion}).  

\begin{figure*}[ht!]
\centering
\subfigure[ $\nu$ = 0.2, a = 0.9, r = 10.01.]
{\includegraphics[width=7.5cm]{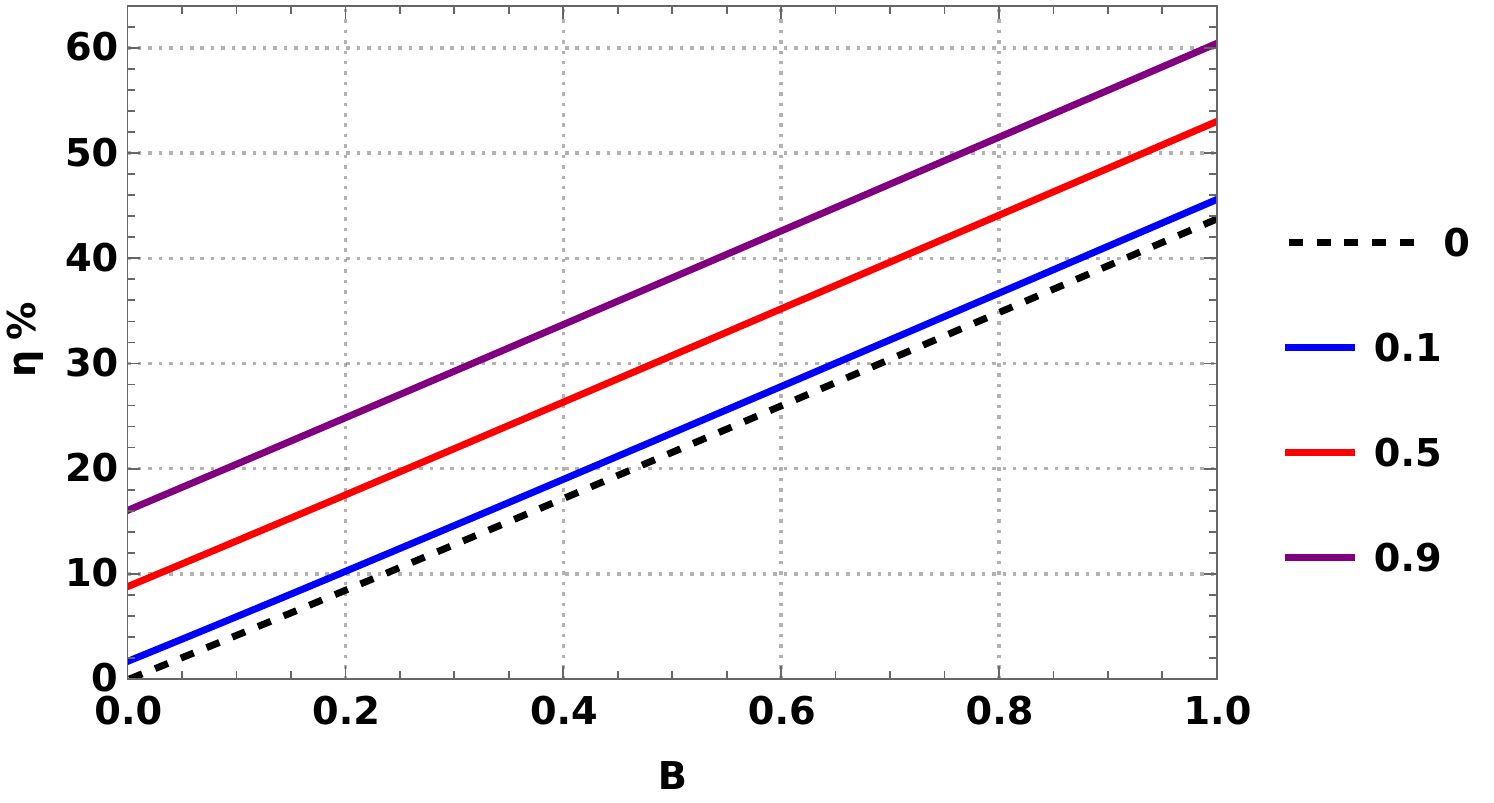}\label{fig:51}}
\hspace{0.4cm}
\subfigure[$\nu$ = 0.9, a = 0.9, r = 2.23. ]
{\includegraphics[width=7.5cm]{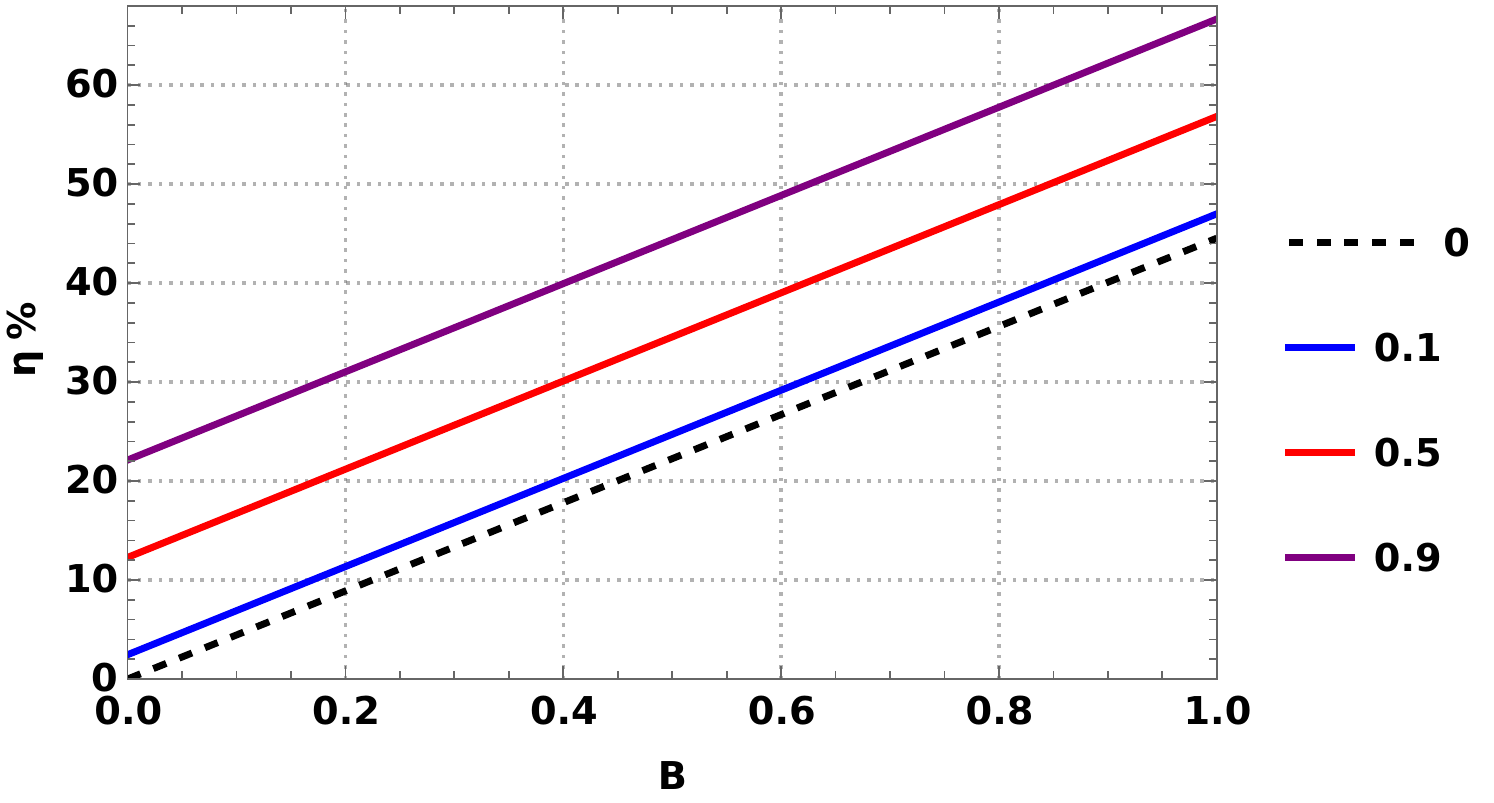}\label{fig:52}}
\hspace{0.4cm}
\subfigure[$\nu$ = 0.2, a = 0.9, B = 0.9. ]
{\includegraphics[width=7.5cm]{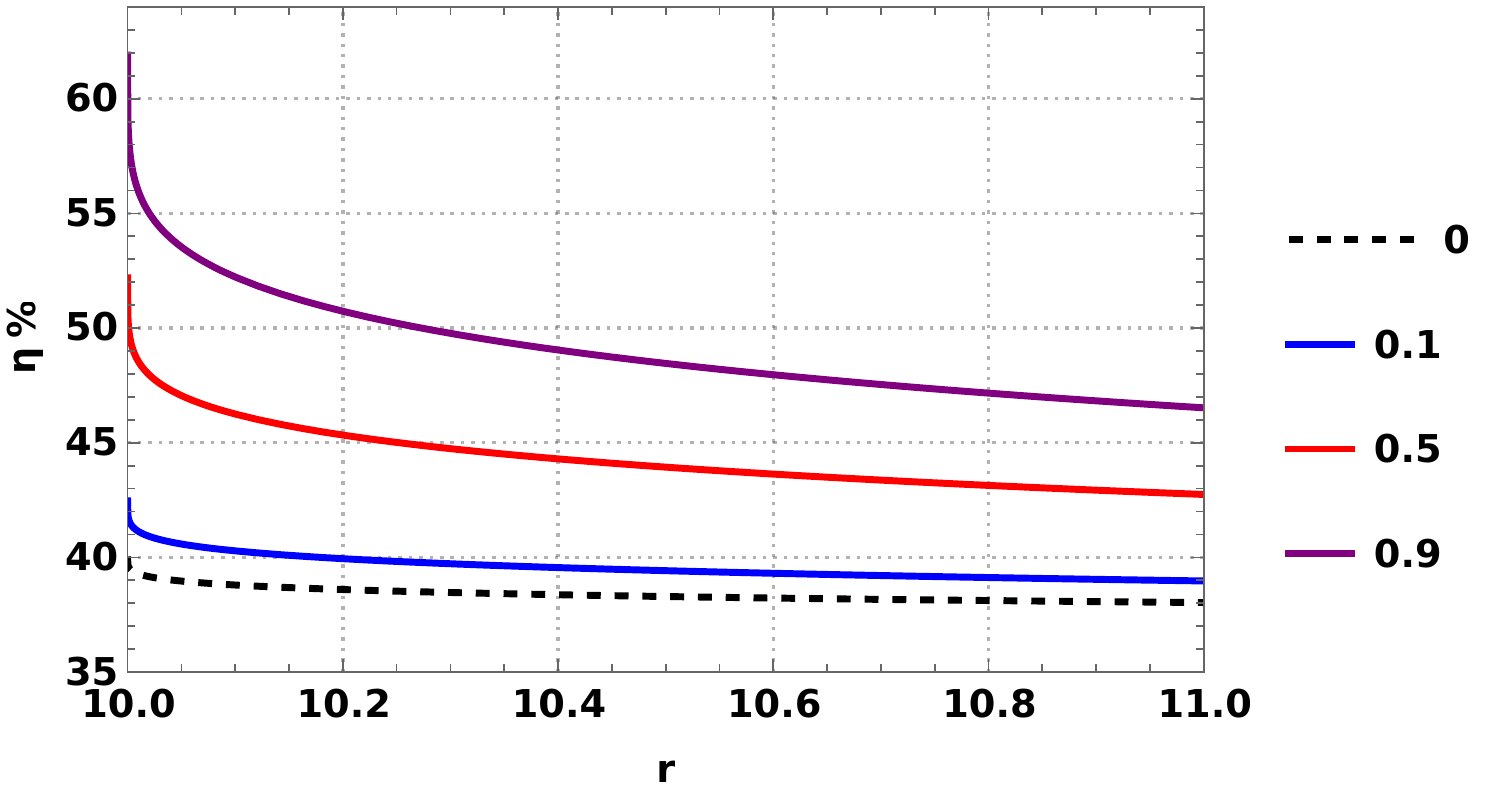}\label{fig:53}}
\hspace{0.4cm}
\subfigure[$\nu$ = 0.9, a = 0.9, B = 0.9. ]
{\includegraphics[width=7.5cm]{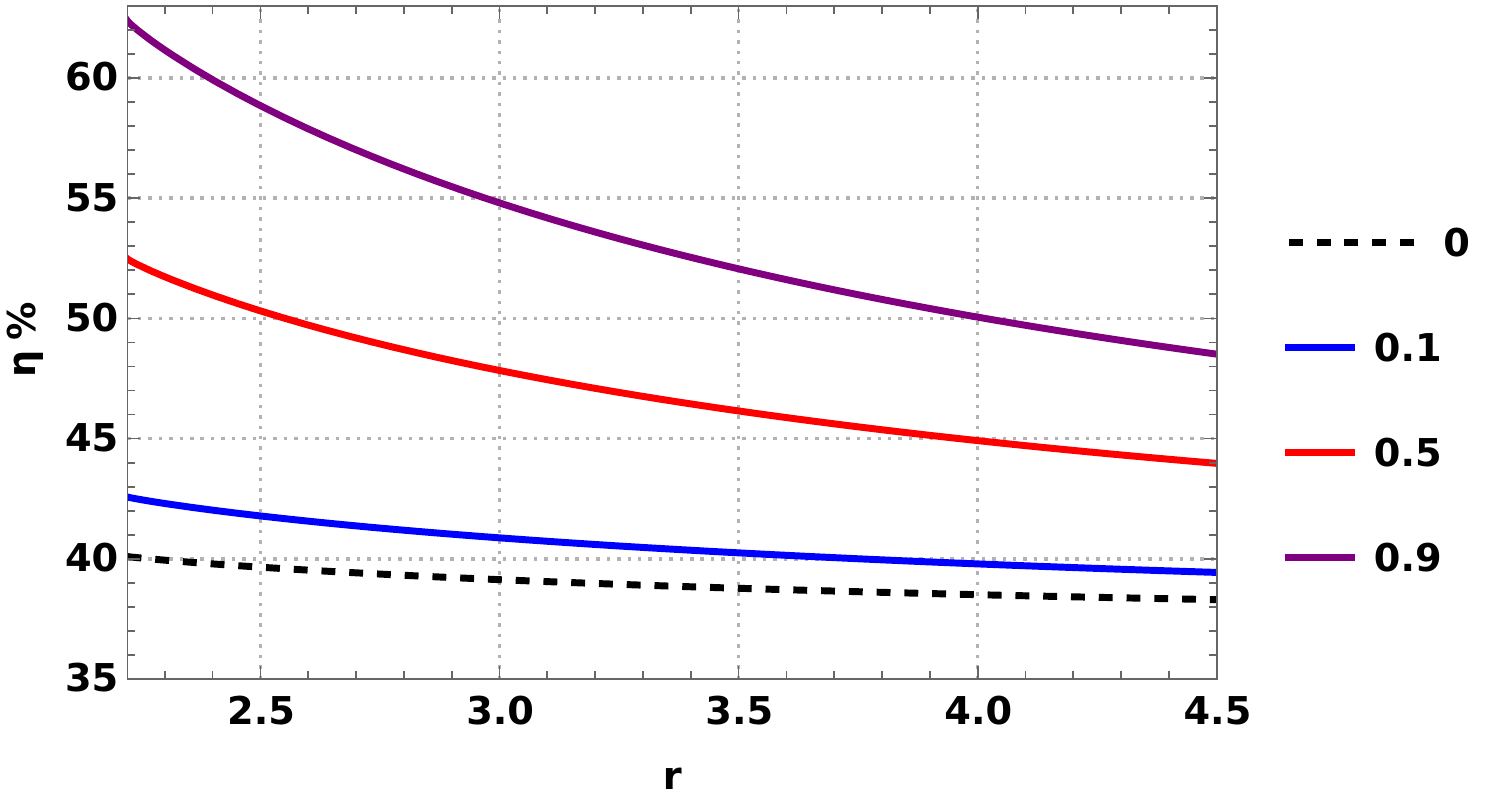}\label{fig:54}}
\hspace{0.4cm}
\subfigure[$\nu$ = 0.5, Q = 0.9, B = 0.5. ]
{\includegraphics[width=7.5cm]{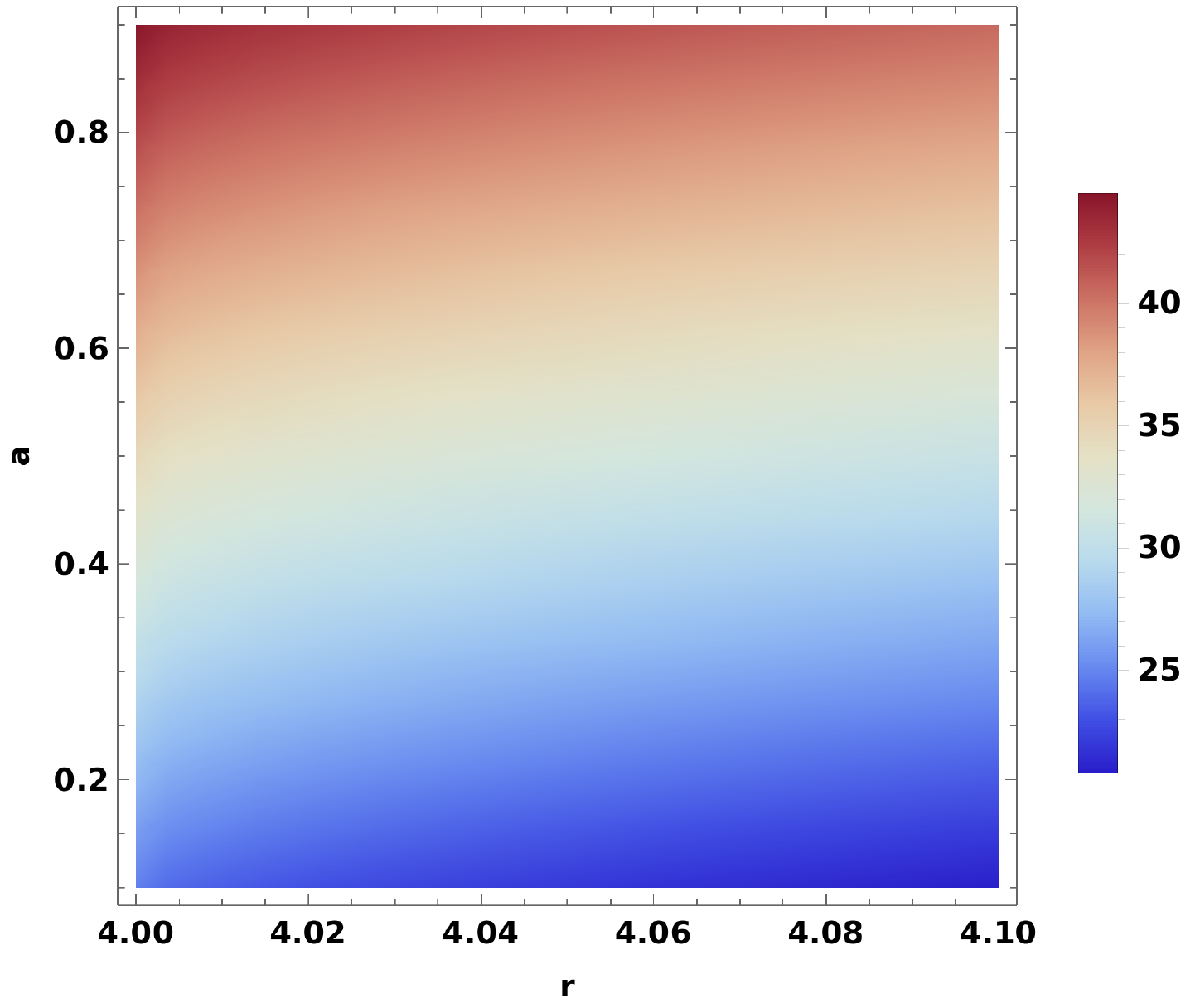}\label{fig:55}}
\hspace{0.4cm}
\subfigure[$\nu$ = 0.9, Q = 0.9, B = 0.9. ]
{\includegraphics[width=7.5cm]{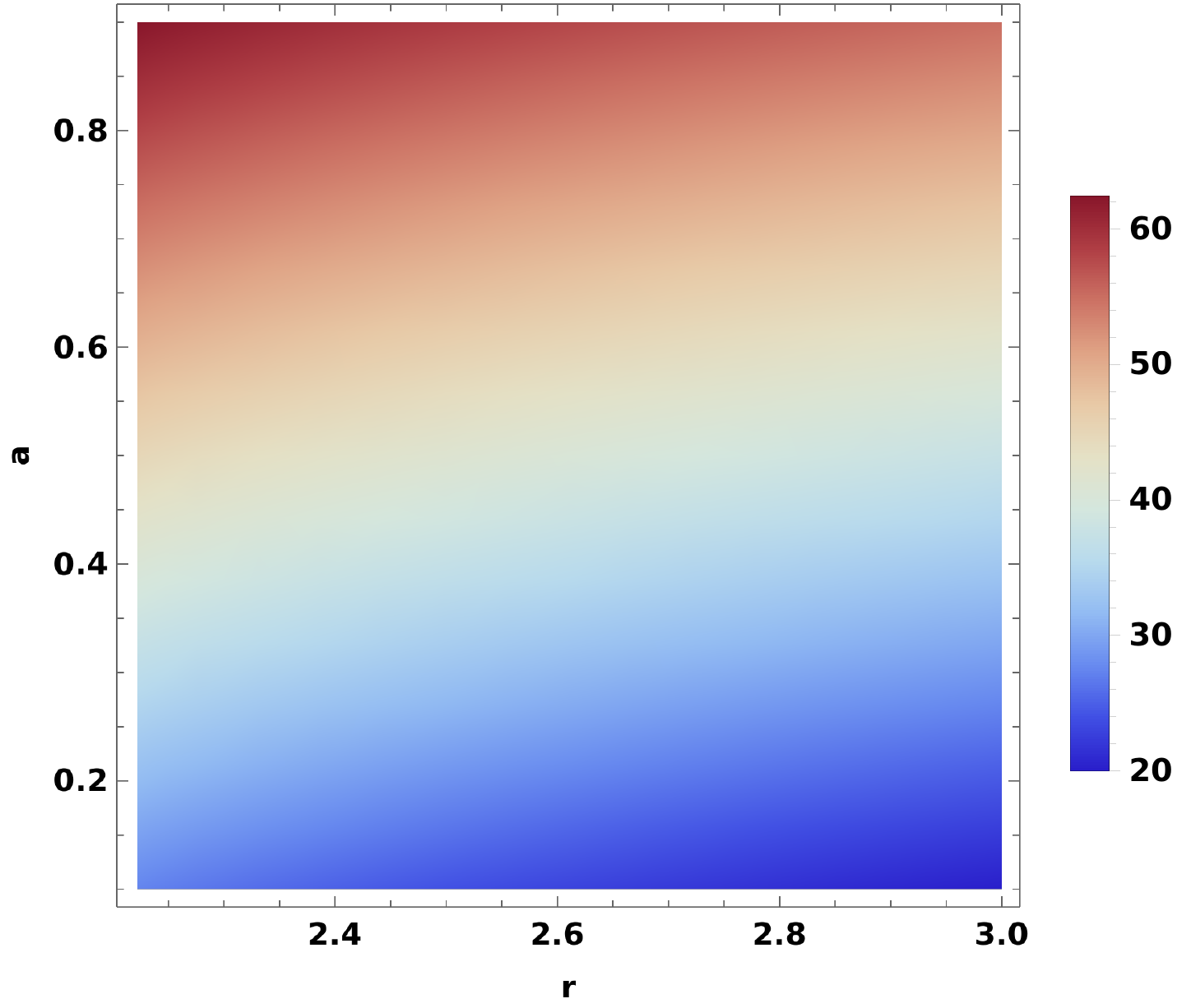}\label{fig:56}}
\hspace{0.4cm}
 \caption{The above Figures show the change in the energy extraction efficiency with respect to the radial distance $r$ (Figs.\,(\ref{fig:51}), (\ref{fig:52})) and the magnetic field intensity parameter $B$ (Figs.\,(\ref{fig:53}), (\ref{fig:54})) in the rotating JNW naked singularity. The different values on the right side of the Figs. (\ref{fig:51}, \ref{fig:52}, \ref{fig:53}, \ref{fig:54}) indicates the variation in an electric charge parameter $(Q)$ around the compact object. The density plots (Figs.\,(\ref{fig:55}), (\ref{fig:56})) indicate the change in energy extraction efficiency with respect to the radius (r) and spin parameter (a). The bar on the right side of the Figs.\,(\ref{fig:55}, \ref{fig:56}) represents the energy extraction efficiency ($\eta$) in percentage. The mass is taken as unity.}
 \label{fig:5}
\end{figure*}

\section{Energy extraction}
\label{sec_energyextra}

In this section, we study the energy extraction mechanism and behavior of energy extraction efficiency for the charged particle around the JNW naked singularity in an asymptotically uniform magnetic field. The mechanism we consider here is the modified version of the Penrose process known as the Magnetic Penrose process, as we are considering the presence of the magnetic field. The magnetic Penrose process is defined in the \cite{Wagh:1985vuj,Tursunov:2019oiq,Wagh:1989zqa}. Hence, here we present a quick overview of the magnetic Penrose process.\\

Let us assume that a charged particle $q_{1}$ with energy $E_{1}$ and mass $m_{1}$ is in the vicinity of the rotating JNW naked singularity which splits into two particles with charge $q_{2}$, energy $E_{2}$, mass $m_{2}$ and particle with charge $q_{3}$, energy $E_{3}$, mass $m_{3}$. Assume that, a particle with charge $q_{3}$ and energy $E_{3}$ enters the negative energy state $E_{3}<0$ and falls into the naked singularity while the other particle with charge $q_{2}$ and energy $E_{2}$ gains this energy and escapes to infinity with $E_{2}>0$ (as the energy is conserved). Taking the conservation laws of different quantities into consideration during this phenomenon, one can write,
$$ E_{(1)} = E_{(2)} + E_{(3)},\hspace{0.8cm} q_{(1)} = q_{(2)} + q_{(3)}, $$
$$ L_{(1)} = L_{(2)} + L_{(3)},\hspace{0.8cm} M_{(1)} = M_{(2)} + M_{(3)}. $$
The canonical momentum is also conserved,
$$ P_{(1)}^{\mu} = P_{(2)}^{\mu} + P_{(3)}^{\mu}. $$
From the conservation law, one can define a relationship for the four velocity $u^{t}$ and $u^{\phi}$ as the angular velocity of the particles,
$$ \Omega = \frac{u^{\phi}}{u^{t}} = \frac{\Dot{\phi}}{\Dot{t}},  $$
and from the conservation $u^{\phi} = \Omega \beta/\alpha$ where, $\alpha$ and $\beta$ are,
$$ \alpha\,=\,\epsilon + \frac{q\,A_{t}}{m},\hspace{0.8cm} \beta\,=\,g_{tt}\,+\,\Omega\,g_{t\phi}, $$
where $\epsilon = E/m$ and by doing the algebraic manipulation one can get the expression for the energy of the escaping particle as,
\begin{equation}
  E_{2}\,=\,\chi\,(E_{1}\,+\,q_{1}\,A_{t})\,-\,q_{2}\,A_{t}, \label{energyeqn} 
\end{equation}
where, $\chi$ is,
$$ \chi\,=\,\frac{\Omega_{1}\,-\,\Omega_{3}}{\Omega_{2}\,-\,\Omega_{3}} \frac{\beta_{2}}{\beta_{1}}, \hspace{0.8cm}  \beta_{i}\,=\,g_{tt}\,+\,\Omega_{i}\,g_{t\phi},$$
where, $\Omega_{i}$ is the angular velocity of the $i^{th}$ particle which can be derived from the geodesic Eq.\,(\ref{geode}). The angular velocity of a different particle is,
\begin{widetext}
 \begin{eqnarray}
      && \Omega_{1}\,=\,\frac{-g_{t\phi} (\pi^{2}_{v}\,+\,g_{tt}) + \sqrt{(\pi^{2}_{v}\,+\,g_{tt}) (g_{t\phi}^{2}\,-\,g_{tt}g_{\phi\phi}) \pi^{2}_{v} }}{\pi^{2}_{v} g_{\phi \phi} + g_{t\phi}^{2}}   ,\\
    && \Omega_{2}\,=\, \frac{-g_{t\phi}\,-\,\sqrt{g_{t\phi}^{2} - g_{tt}g_{\phi\phi}} }{g_{\phi\phi}}   ,\\
    && \Omega_{3}\,=\,\frac{-g_{t\phi}\,+\,\sqrt{g_{t\phi}^{2} - g_{tt}g_{\phi\phi}} }{g_{\phi\phi}}   ,
 \end{eqnarray}   
\end{widetext}
here, $\pi_{v}\,=\,-(\epsilon\,+\,q\,A_{t}/m)$. Using the usual definition of efficiency,
$$ \eta\,=\, \frac{E_{3}\,-\,E_{1}}{E_{1}} = \frac{-E_{2}}{E_{1}},  $$
using the above expression and by doing algebraic manipulation with Eq.\,(\ref{energyeqn}), we get the efficiency of the magnetic Penrose process in the form,
\begin{equation}
    \eta\,=\, \frac{\chi\,q_{1}\,A_{t}\,-\,q_{3}\,A_{t}}{E_{1}} + \chi\,-\,1. \label{effi}
\end{equation}
The energy extraction efficiency in the magnetic Penrose process is reduced to the original Penrose process in the absence of an electromagnetic field. Figs.\,(\ref{fig:5}) represent the change in efficiency with respect to the radial distance $r$ (Figs.\,(\ref{fig:51}), (\ref{fig:52})) and the magnetic field intensity $B$ (Figs.\,(\ref{fig:53}), (\ref{fig:54})) with different electric charge (Q). The detailed discussion is given in section (\ref{sec_conclusion}).

\section{Discussion and Conclusions}
\label{sec_conclusion}
In this paper, we investigate the energy extraction phenomena from a rotating JNW naked singularity case. For this, we considered rotating JNW naked singularity in a uniform magnetic field. We briefly discussed the importance of negative energy orbits in energy extraction and considered the negative energy states for a charged particle moving in this geometry. In order to understand the magnetic Penrose process we considered the splitting of a charged particle in JNW naked singularity surrounded by an electromagnetic field. The following are the outcomes of this study:

\begin{itemize}
    \item In \cite{Misner:1973prb}, it is given that for in-falling neutral particle for Kerr black hole, the negative energy orbits exist inside the ergoregion. Whereas, for in-falling charged particle, the region of the negative energy orbits is larger than the ergoregion. This region is known as effective ergoregion. For energy extraction previously, it was shown by Ruffini that the existence of negative energy orbits is necessary. Apart from that, it is known and in \cite{Karmakar:2017lho} authors show that there does not exist ergoregion in the rotating JNW naked singularity. However, in our study, using the uniform magnetic field around the rotating JNW naked singularity, we show the presence of negative energy orbits for the charged particles. Hence, one can say that there exists an effective ergoregion in the rotating JNW naked singularity.

    \item We proved that the negative energy orbits exist for $0.1\,<\,\nu \,<\,1$ with the conditions $E\,<\,0$, $L\,<\,0$ and the charge of moving particle (q) and induced electromagnetic charge (Q) in the geometry are such that $q\,Q\,<\,0$ in the rotating JNW naked singularity surrounded by the uniform electromagnetic field. Figs.\,(\ref{fig:1}) shows the change in the energy (E) with respect to the radial distance (r) for different parameters. The region of the negative energy orbits increases as the value of the parameter $\nu$ is increased and also with the increasing magnetic field intensity parameter ($B$) and spin parameter ($a$). The reason behind this behavior is, with the increasing value of the magnetic field intensity parameter and spin parameter, the required energy for the particle to escape to infinity also increases as discussed in the section (\ref{sec_neo}). The Figs.\,(\ref{fig:1}) of negative energy orbits for $\nu=1$ resemble the plots of negative energy orbits for the Kerr geometry cases. 

    \item One may see that in the absence of the induced electric charge, the negative energy orbits are exists from Figs.\,(\ref{fig:14}, \ref{fig:17}, \ref{fig:110}). Apart from that, when the magnetic field reduces to zero then there is an absence of the negative energy orbits. This concludes that negative energy orbits exist in the rotating JNW geometry in the presence of a magnetic field or electromagnetic field. The induced electric charge might play a crucial role in the higher negative energy orbit region or energy extraction efficiency but is not necessary for the existence of the negative energy orbit or the energy extraction.   
   
   \item In general, when the region of the negative energy orbits increases because of a change in the value of the magnetic field intensity parameter, spin parameter, or the metric parameter $\nu$, the energy extraction efficiency also increases. This is because the escaping particle gets more energy as the region of negative energy orbits increases. This is also observed in Figs.\,(\ref{fig:5}) as a result of an investigation. The result of increasing energy extraction efficiency was expected as the region of the negative energy orbits was also increasing.

   \item The Figs.\,((\ref{fig:51}), (\ref{fig:52})) indicates the decreasing rate of the energy extraction efficiency with respect to the radial distance (r) as one moves away from the singularity. Figs.\,(\ref{fig:53}), (\ref{fig:54})) show that energy extraction efficiency increases with increasing value of magnetic field intensity, the spin parameter. This behaviour is because of the increasing negative energy orbits region in the presence of magnetic field and rotation as escaping particle gets more energy. The energy extraction efficiency also increases with the increasing value of the metric parameter $\nu$. Figs.\,(\ref{fig:55}, \ref{fig:56}) shows the change in energy extraction efficiency rate with respect to the spin parameter and radial distance. As expected it indicates that for the higher spin parameter, the efficiency rate is increased compared to the lower spin parameter.

   \item  The energy extraction efficiency for rotating JNW naked singularity case is 60\% for metric parameter $\nu=0.9$, magnetic field intensity $B=0.9$, and spin parameter $a=1$. While for the Kerr black hole and Kerr naked singularity case, the efficiency is around $20.7\%$ and $150\%$ respectively. There are many possible reasons why there exists a high-efficiency rate in the Kerr naked singularity than rotating JNW naked singularity. The first reason is that a spin parameter is higher in the Kerr naked singularity $(a>M)$ with the existence of an ergoregion. While in JNW spacetime, there is no ergoregion but negative energy orbits are present in so-called effective ergoregion, therefore the efficiency of energy extraction from the Kerr naked singularity is higher than JNW naked singularity case. Moreover, we may have to look further for the relation between the presence of an ergoregion and negative energy orbits. These fundamental questions need to be explored critically in future.

\end{itemize}

\acknowledgments
 The authors would like to express their gratitude towards Dr. Daniela Pugliese for her valuable suggestions and comments. V.P. would like to acknowledge the support of the SHODH fellowship (ScHeme Of Developing High quality research—MYSY).

\appendix

\section{}
\label{sec_app}
The expression for the energy states can be written as,
\begin{widetext}
    \begin{eqnarray}
         E\,=\,\frac{f_1+f_2-\left(f_3\pm \sqrt{\left(f_4-f_5+f_6\right){}^2-4 f_7 \left(f_8+f_9+f_{10}-f_{11}+f_{12}-f_{13}-f_{14}-f_{15}\right)}\right)}{f_0},
    \end{eqnarray}
\end{widetext}
where, $f_{i}$\,($i=0,,,15$) is function of $[B, m, q, Q, \nu, a, L, r]$,
\begin{widetext}
\begin{eqnarray}
&& f_{0}\,=\, 2 \left(8 a^2 \nu ^2 \left(1-\frac{2 m}{\nu  r}\right)^{\nu }-4 a^2 \nu ^2 \left(1-\frac{2 m}{\nu  r}\right)^{2 \nu }-8 m \nu  r+4 \nu ^2 r^2\right),\\
&& f_{1}\,=\,-16 a B m \nu  q r \left(1-\frac{2 m}{\nu  r}\right)^{\nu }+8 m \nu  q Q r \left(1-\frac{2 m}{\nu  r}\right)^{\nu }-8 m \nu  q Q r+4 \nu ^2 q Q r^2,\\
&& f_{2}\,=\, 8 a^3 B \nu ^2 q \left(1-\frac{2 m}{\nu  r}\right)^{\nu }+4 a^2 \nu ^2 q Q \left(1-\frac{2 m}{\nu  r}\right)^{\nu }+8 a B \nu ^2 q r^2 \left(1-\frac{2 m}{\nu  r}\right)^{\nu
   }+8 a L \nu ^2 \left(1-\frac{2 m}{\nu  r}\right)^{\nu },\\
&& f_{3}\,=\, -4 a^2 \nu ^2 q Q \left(1-\frac{2 m}{\nu  r}\right)^{2 \nu }-8 a L \nu ^2 \left(1-\frac{2 m}{\nu  r}\right)^{2 \nu }+4 \nu ^2 q Q r^2 \left(1-\frac{2 m}{\nu  r}\right)^{\nu },\\
&& f_{4}\,=\, 16 a B m \nu  q r \left(1-\frac{2 m}{\nu  r}\right)^{\nu }-8 m \nu  q Q r \left(1-\frac{2 m}{\nu  r}\right)^{\nu }+8 m \nu  q Q r-4 \nu ^2 q Q r^2,\\
&& f_{5}\,=\, -8 a^3 B \nu ^2 q \left(1-\frac{2 m}{\nu  r}\right)^{\nu }-4 a^2 \nu ^2 q Q \left(1-\frac{2 m}{\nu  r}\right)^{\nu }-8 a B \nu ^2 q r^2 \left(1-\frac{2 m}{\nu  r}\right)^{\nu
   }+8 a L \nu ^2 \left(1-\frac{2 m}{\nu  r}\right)^{\nu },\\
&& f_{6}\,=\, 4 a^2 \nu ^2 q Q \left(1-\frac{2 m}{\nu  r}\right)^{2 \nu }+8 a L \nu ^2 \left(1-\frac{2 m}{\nu  r}\right)^{2 \nu }+4 \nu ^2 q Q r^2 \left(1-\frac{2 m}{\nu  r}\right)^{\nu },\\
&& f_{7}\,=\, 8 a^2 \nu ^2 \left(1-\frac{2 m}{\nu  r}\right)^{\nu }-4 a^2 \nu ^2 \left(1-\frac{2 m}{\nu  r}\right)^{2 \nu }-8 m \nu  r+4 \nu ^2 r^2,\\
&& f_{8}\,=\, 2 a^2 B^2 m \nu  q^2 r-4 B^2 m^2 q^2 r^2+4 B^2 m \nu  q^2 r^3-2 m \nu  q^2 Q^2 r,\\
&& f_{9}\,=\, -4 a^2 B^2 m \nu  q^2 r \left(1-\frac{2 m}{\nu  r}\right)^{\nu }-4 a B m \nu  q^2 Q r \left(1-\frac{2 m}{\nu  r}\right)^{\nu }-8 B L m \nu  q r \left(1-\frac{2 m}{\nu 
   r}\right)^{\nu }+8 m \nu  r \left(1-\frac{2 m}{\nu  r}\right)^{\nu },\\
&& f_{10}\,=\, -2 a^2 B^2 m \nu  q^2 r \left(1-\frac{2 m}{\nu  r}\right)^{2 \nu }+4 a B m \nu  q^2 Q r \left(1-\frac{2 m}{\nu  r}\right)^{2 \nu }+4 m \nu  q^2 Q^2 r \left(1-\frac{2 m}{\nu 
   r}\right)^{\nu }-2 m \nu  q^2 Q^2 r \left(1-\frac{2 m}{\nu  r}\right)^{2 \nu },\\
&& f_{12}\,=\, 2 a^4 B^2 \nu^2 q^2 \left(1-\frac{2 m}{\nu  r}\right)^{\nu }+2 a^3 B \nu ^2 q^2 Q \left(1-\frac{2 m}{\nu  r}\right)^{\nu }+4 a^2 B L \nu ^2 q \left(1-\frac{2 m}{\nu 
   r}\right)^{\nu }+4 a L \nu ^2 q Q \left(1-\frac{2 m}{\nu  r}\right)^{\nu },\\
&& f_{13}\,=\, 2 a^2 B^2 \nu ^2 q^2 r^2 \left(1-\frac{2 m}{\nu  r}\right)^{\nu }+2 a B \nu ^2 q^2 Q r^2 \left(1-\frac{2 m}{\nu  r}\right)^{\nu }+4 B L \nu ^2 q r^2 \left(1-\frac{2 m}{\nu 
   r}\right)^{\nu }+4 \nu ^2 r^2 \left(1-\frac{2 m}{\nu  r}\right)^{\nu },\\
&& f_{14}\,=\, a^4 B^2 \nu ^2 q^2 \left(1-\frac{2 m}{\nu  r}\right)^{2 \nu }-4 a L \nu ^2 q Q \left(1-\frac{2 m}{\nu  r}\right)^{2 \nu }-4 L^2 \nu ^2 \left(1-\frac{2 m}{\nu  r}\right)^{2
   \nu }+2 \nu ^2 q^2 Q^2 r^2 \left(1-\frac{2 m}{\nu  r}\right)^{\nu },\\
&& f_{15}\,=\, 2 a^3 B \nu ^2 q^2 Q \left(1-\frac{2 m}{\nu  r}\right)^{2 \nu }+a^2 B^2 \nu ^2 q^2 r^2 \left(1-\frac{2 m}{\nu  r}\right)^{2 \nu }-2 a B \nu ^2 q^2 Q r^2 \left(1-\frac{2
   M}{\nu  r}\right)^{2 \nu }+\nu ^2 q^2 Q^2 r^2 \left(1-\frac{2 m}{\nu  r}\right)^{2 \nu }.
\end{eqnarray}
\end{widetext}

\end{document}